\documentclass[12pt]{iopart}
\usepackage{iopams}
\usepackage[latin1]{inputenc}
\usepackage{graphicx}
\usepackage{times}
\usepackage{bm}

\begin{document}

\title[Detecting FF superconductors by an Andreev interferometer]
{Detecting Fulde-Ferrell superconductors by an Andreev interferometer}

\author{Wei Chen$^{1,2}$, Ming Gong$^{3*}$, R. Shen$^{1\dagger}$, D. Y. Xing$^{1}$}

\address{1 National Laboratory of Solid State Microstructures and Department of Physics, Nanjing University, Nanjing 210093, China\\
2 College of Science, Nanjing University of Aeronautics and Astronautics, Nanjing 210016, China\\
3 Department of Physics and Centre for Quantum Coherence, The Chinese University of Hong Kong, Shatin, N.T., Hong Kong, China\\

* E-mail: skylark.gong@gmail.com\\
$\dagger$ E-mail: shen@nju.edu.cn}

%\homepage{http://www.fotonik.dtu.dk/quantumphotonics}

\begin{abstract} %max 200 words (present 164)
We propose an Andreev interferometer, based on a branched Y-junction, to detect the finite momentum pairing in Fulde-Ferrell (FF) superconductors. In this interferometer, the oscillation of subgap conductance is a unique function of phase difference between the two channels of the Y-junction, which is determined by the phase modulation of the order parameter in the FF superconductors. This interferometer has the potential not only to determine the magnitude but also the direction of the momentum of Cooper pairs in the FF superconductor. The possible applications of the interferometer in the identification of the finite momentum pairing in non-centrosymmetric superconductors are also discussed.
\end{abstract}

\maketitle
%\tableofcontents
%42.25.Dd Wave propagation in random media
%42.50.-p	Quantum optics
%78.67.Hc	Quantum dots

\section{Introduction}

Cooper pair can carry finite total momentum. Just after the celebrated Bardeen-Cooper-Schrieffer (BCS) theory of superconductivity \cite{BCS},  in which the Cooper pair is formed by two fermions with opposite momenta, Fulde and Ferrell (FF) predicted that finite momentum pairing may occur in some type-II superconductors at strong magnetic field \cite{Fulde}, where the order parameter $\Delta_{FF} = \Delta e^{i{\bm Q}\cdot {\bm x}}$, with ${\bm Q}$ the total momentum of the Cooper pair and $\Delta$ a constant. It corresponds to pairing between ${\bm k}$ and $-{\bm k}+{\bm Q}$ in the momentum space. A similar state, which may have lower energy than the FF phase, was predicted by Larkin and Ovchinnikov (LO) \cite{Larkin}, with $\Delta_{LO} = \Delta \cos({\bm Q}\cdot {\bm x})$. The FFLO state admits the coexistence of magnetism and superconductivity and is a key concept to understand the superconducting behavior in some type-II superconductors, e.g., layered \cite{Croitoru},  heavy-fermion \cite{Radovan,Yuji,Gloos,Bianchi,Kenzelmann} and organic \cite{Singleton, Lortz} superconductors. In the past five decades, great endeavors have been paid trying to unveil this novel phase, unfortunately, only indirect evidences related to the possible finite momentum pairing have been reported. The basic reason is that the reported experimental tools, including Andreev reflection (AR) \cite{Kontos,Park,Park2,Krawiec}, specific heat \cite{Radovan,Bianchi,Bianchi1,Movshovich}, nuclear magnetic resonance \cite{Kakuyanagi} and ultrasound velocities \cite{Ultrasound}, only measure some anomalous properties of the superconductors, which may be caused by other phase transitions \cite{Kun}.

Directly detecting the Cooper pair momentum, without doubt, provides the most convincing evidence for finite momentum pairing. In this paper, we propose an Andreev interferometer based on a branched Y-junction for this particular purpose. In our interferometer, the subgap conductance oscillation is uniquely determined by the phase modulation of the order parameter, thus provides a distinctive method to detect the Cooper pair momentum in FF superconductors. The device is very robust because all the uncontrollable phases during the multiple scatterings at the Y-junction are exactly canceled out. This interferometer may have intriguing applications in non-centrosymmetric superconductors \cite{Yokoyama}, including Li$_2$(Pd$_{x}$Pt$_{1-x}$)$_3$B \cite{Yuan, Lee}, CePt$_3$Si \cite{Frigeri},  CeRh(Ir)Si$_3$ \cite{Tada},  CeCoIn$_5$/YbCoIn$_5$ superlattice \cite{Swee} and SrTiO$_3$/LaAlO$_3$ interface \cite{Patrick} {\it etc.}. Generally, the interplay between the spin-orbit coupling (induced by either the bulk or the structure inversion symmetry breaking) and the in-plane Zeeman field not only stabilizes the FF phase against the formation of the LO phase \cite{Dimitrova1,Samokhin,Kaur1,Kaur2,Dimitrova2}, but also greatly enlarges its phase size in the parameter space \cite{ZZ1,ZZ2}. In this sense, our interferometer is best suitable to facilitate the identification of the possible finite momentum pairing in these materials.

\begin{figure}
\centering
\includegraphics[width=0.8\textwidth]{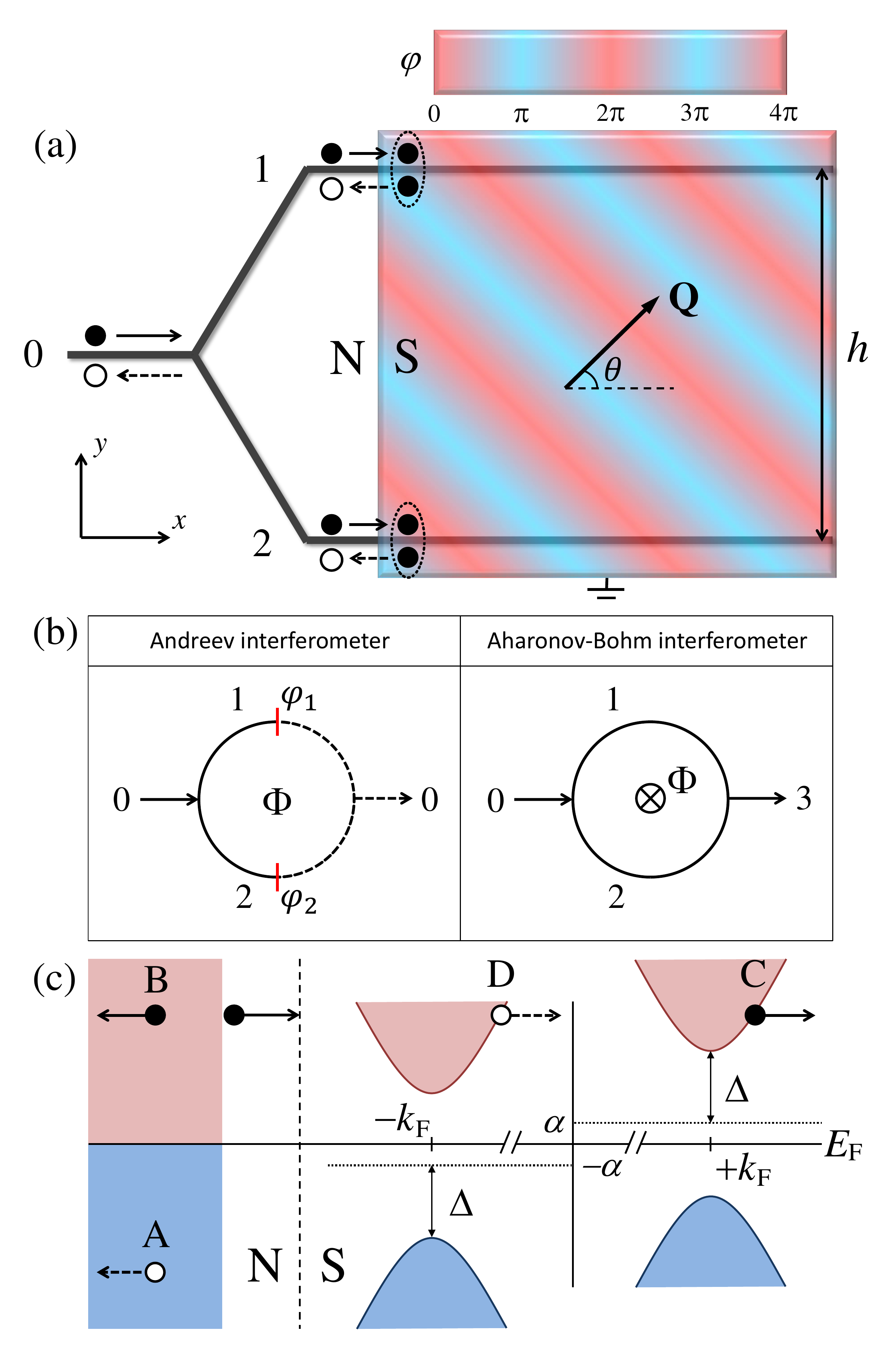}
\caption{(color online) Andreev interferometer to detect the Cooper pair momentum. (a) A branched Y-junction in proximity to a two dimensional FF superconductor. The phase modulation of the order parameter along the $\theta$-direction is sketched by the color bar. The electron and the Andreev reflected hole are sketched by the filled and open circles, with their propagations represented by the solid and dashed lines, respectively. (b) Analogy between an Andreev interferometer and an Aharonov-Bohm interferometer. (c) Schematics of the dispersion of the FF superconductor and the scattering processes at the NS interfaces. The quasiparticle excitation spectra near the Fermi wave vectors ($\pm k_{F}$) are shifted by the values of $\pm\alpha$. Processes A, B, C and D denote the AR, the normal reflection, and the electron-like and hole-like quasiparticle transmissions, respectively.} \label{fig1}
\end{figure}

\section{Model and Basic Idea}

The periodic phase modulation in the order parameter is essential to construct the Andreev interferometer, as shown in Fig. \ref{fig1}(a). We consider a nanowire Y-junction in proximity to a two dimensional FF superconductor, which introduces a finite pair potential in the two channels. To be specific, the two channels are set along the $x$-direction, with a spatial separation of $h$ in the $y$-direction. The pairing in the FF superconductor occurs with a finite momentum $\bm{Q}$ = $(Q_{x}, Q_{y})$, and the order parameter takes the form of $\Delta e^{i{\bm Q}\cdot {\bm x}}$ for the $s$-wave pairing. In this work, ${\bm Q}$ is assumed to be changed {\it in situ} by an external magnetic field. The proposed interferometer still works with the unconventional $d$-wave pairing. This is because the projection of the order parameter of the FF superconductor into one dimensional channels eliminates the effect of the internal phase difference in $d$-wave superconductors. In other words, the underlying mechanism for FF pairing is not essential in our interferometer.

The basic idea is as follows. An electron injected from channel $0$ of the Y-junction is split into channels $1$ and $2$. For a transparent normal metal-superconductor (NS) interface, the incident electron penetrates into the superconductor and forms a Cooper pair with another electron, leaving back an Andreev reflected hole, which acquires an extra phase equal to the macroscopic phase of the superconductor \cite{Spivak,Nakano}. Thus, the phase difference $\Phi = Q_{y}h$ between the two channels due to the phase modulation in the FF superconductor can be registered accurately by the reflected hole, leading to the interference when the hole paths finally get combined at the left channel $0$. As a result, a total AR probability proportional to $1+\cos\Phi$ is expected for a symmetric Y-junction. It is interesting to note that the picture is resemblant to the well-known Aharonov-Bohm interferometer, if we sketch the hole paths as a mirror symmetry of the electron paths about the NS interfaces in Fig. \ref{fig1}(b). In the practical case, channels 1 and 2 may have different lengths, there are multiple scatterings at the Y-junction, and barriers may exist at the NS interfaces. These effects may complicate the interference between the two channels, however, as shown below, the basic idea and qualitative results still hold.

\section{Theory and Results}

In order to find the total AR amplitude in channel 0, we first need to find the scattering coefficients at the NS interfaces and then combine them with the scattering matrix at the Y-junction. In Fig. \ref{fig1}(a), the Y-junction is located at $x=0$ and the NS interface at $x=L$. The nanowire can be described by the Bogoliubov-de Gennes (BdG) equation,
\begin{equation}\label{BdG}
\left(
  \begin{array}{cc}
    h_{j}(x) & \Delta_{j}(x) \\
\Delta_{j}^*(x) & -h_{j}(x) \\
  \end{array}
\right)\left(
         \begin{array}{cc}
           u_{j} \\ v_{j} \\
         \end{array}
       \right)
=E\left(
         \begin{array}{cc}
           u_{j} \\ v_{j} \\
         \end{array}
       \right),
\end{equation}
where $h_{j}(x)=-\hbar^2/(2m)\partial_{x}^{2}-\mu+U_{j}(x)$ defines the free electron in channel $j$ ($j=1,2$) with a chemical potential $\mu$ and a barrier $U_{j}(x)=U_{j}\delta(x-L)$ at the NS interface. The effects due to the Land\'e $g$-factor and the spin-orbit coupling are assumed to be negligible in the nanowires,
so that the BdG equation is written in a $2\times 2$ form with the external magnetic field being absent, which although is necessary for the FF superconductor.
Due to the proximity effect with an FF superconductor, a finite pair potential is induced in the nanowire, which is represented by the step function $\Delta_{j}(x)=\Delta e^{i(Q_{x}x+\varphi_{j})}\Theta(x-L)$ with the phase $\varphi_{j} = Q_{y}y_{j}$ and $y_j$ being the vertical ordinate of the channel $j$. A standard Green function approach shows that, although the induced pairing strength $\Delta$ may be smaller than that in the bulk, the phase modulation of the order parameter in the FF superconductor is always accurately registered by the induced pair potential in the nanowire (See appendix 1 for details).

The BdG equation (\ref{BdG}) can be solved by a Galilean transformation $u_{j}\rightarrow u_{j} e^{i Q_{x}x/2}$ and $v_{j}\rightarrow v_{j}e^{-i Q_{x} x/2}$, and the quasiparticle excitation spectra around the Fermi wave vectors ($\pm k_{F}$) are given by
\begin{equation}\label{ei}
E_{\pm}(q)=\pm\alpha+\sqrt{(\hbar v_{F}q)^2+\Delta^2},
\end{equation}
where $q$ is a small wave vector measured from $\pm k_{F}$, $v_{F}$ is the Fermi velocity, and the conditions $Q_{x}\ll k_{F}$ and $\Delta\ll\mu$ have already been taken into account. The quasiparticle energy around $\pm k_{F}$ is shifted by a value of $\alpha=\hbar v_{F}Q_{x}/2$ due to the Cooper pair momentum, which is sketched in Fig. \ref{fig1}(c). We find that the $x$-component of the Cooper pair momentum results in an energy split at $\pm k_{F}$, while the $y$-component contributes to a phase difference $\Phi$ between the two channels.

The scattering amplitudes at the NS interface can be obtained by the Blonder-Tinkham-Klapwijk (BTK) approach \cite{Blonder}. By matching the wave functions at the NS interface, the AR amplitude for an incident electron is obtained as $a_{j}=u_{0}^{-}v_{0}^{+}e^{-i\varphi_{j}}/\gamma_{j}$, where $\gamma_{j}=u_{0}^{+}u_{0}^{-}(1+Z_{j}^{2})-v_{0}^{+}v_{0}^{-}Z_{j}^{2}$, $Z_{j}=mU_{j}/(\hbar^{2}k_{F})$ is the dimensionless barrier strength, and $u_{0}^{\pm}=\sqrt{[1+\sqrt{(E\mp\alpha)^{2}-\Delta^{2}}/(E\mp\alpha)]/2}$ and $v_{0}^{\pm}=\sqrt{1-(u_{0}^{\pm})^{2}}$ are the electron and hole components of the wave functions around $\pm k_{F}$, respectively. The AR amplitude for an incident hole is denoted as $a_{j}'$ and can be obtained similarly (See appendix 2 for details). Since the incident electron and the Andreev reflected hole move with the same momentum but opposite directions, the phases accumulated by them during the propagation are canceled out exactly. Consequently, the AR amplitudes $a_{j}$ and $a_{j}'$ only depend on the macroscopic phase of the superconductor ($\varphi_{j}$) and have nothing to do with the channel lengths. The phase difference $\varphi_{1}-\varphi_{2}=\Phi$ is defined by the Cooper pair momentum $Q_{y}$.

A symmetric Y-junction is generally favorable to observe the strongest interference effect. Such that the scattering amplitudes for the electron at the Y-junction can be fully parameterized as follows \cite{Buttiker}: $t$ for the transmission between channel 0 and channel $j$, $\tau$ for the transmission between channel 1 and channel 2, and $\rho$ for the reflection within channel $j$. Since there is no AR process at the Y-junction, the energy dependence of $t$, $\tau$, and $\rho$ in scale of $\Delta$ can be neglected. The scattering amplitudes for the hole are just the complex conjugate of those for the electron due to the particle-hole symmetry of the BdG equation.

\begin{figure}
\centering
\includegraphics[width=0.8\textwidth]{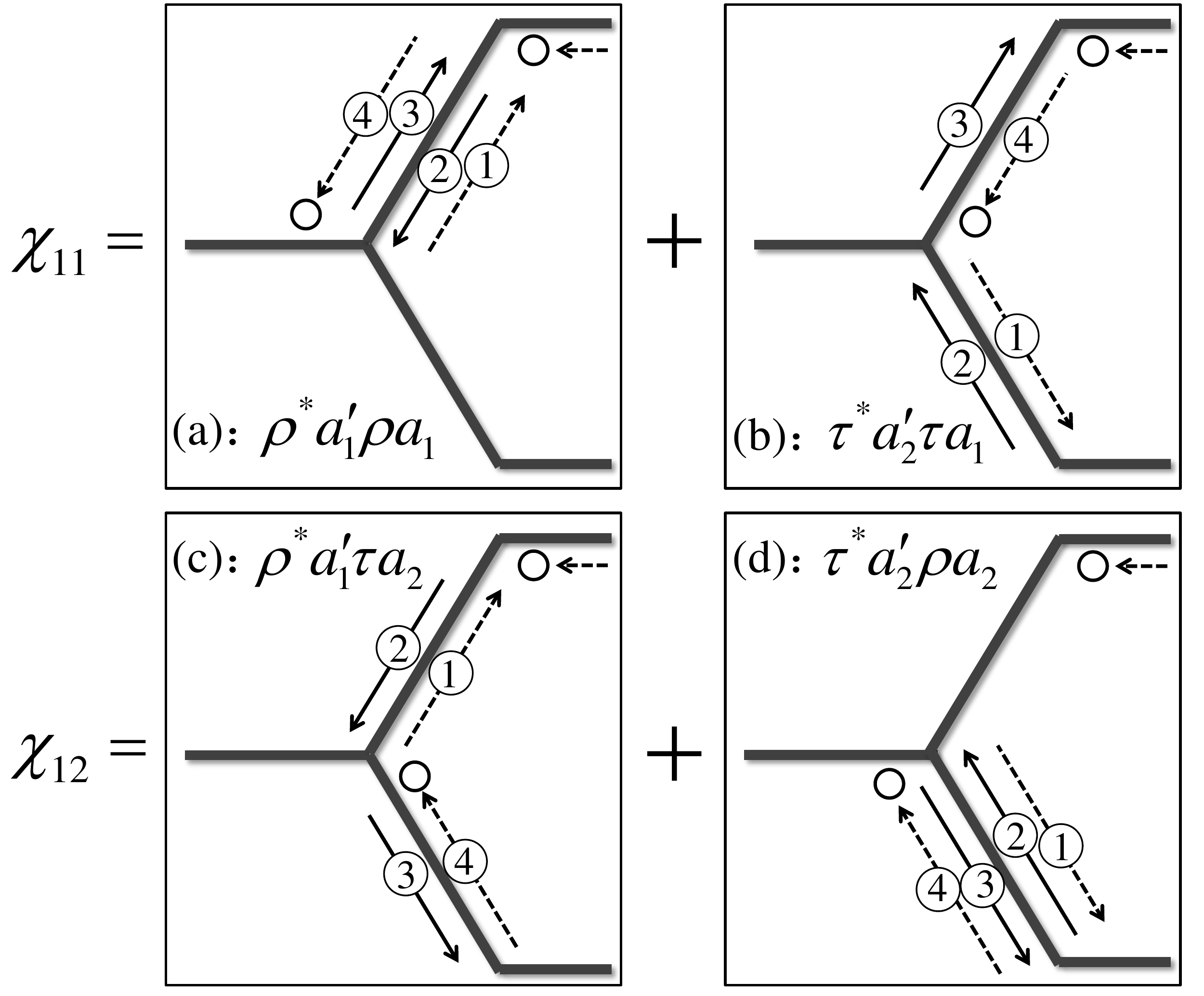}
\caption{Basic scattering loops between the Y-junction and the NS interfaces. The solid and dashed lines with arrows denote the electron and hole propagations, respectively. The number in the circle represents the sequence of the scattering. The circles without number are the Andreev reflected hole before and after the scattering loop. The scattering amplitudes in each loop are labeled correspondingly.}
\label{fig2}
\end{figure}

When a bias voltage $eV$ is utilized in channel 0 and the superconductor is grounded, the incident electron is multiply scattered between the Y-junction and the NS interfaces before it finally goes back into channel 0 as an electron or a hole. The total AR amplitude $\tilde{a}$ in channel 0 can be obtained by combining the scattering amplitudes at the NS interfaces and those at the Y-junction in a standard scattering matrix approach (See appendix 3 for details). We note that, since the NS interface here is actually consisted of the same nanowire and the AR signal is strongest when the NS interface is transparent, the limiting case of $Z_{j}=0$ is particularly important and practically useful. In this simple case, the combination of the scattering coefficients can be fulfilled in a physically transparent way by a Feynman path summation.

The basic scattering loops between the Y-junction and the NS interfaces are sketched in Fig. \ref{fig2}, where $\chi_{ij}$ represents the scattering amplitude for a backward hole (moving towards $x=0$) in channel $i$ scattered into a backward hole in channel $j$. From Fig. \ref{fig2}, we obtain $\chi_{11} =(\rho^{*}a_{1}'\rho+\tau^{*}a_{2}'\tau)a_{1}$ and $\chi_{12} =(\rho^{*}a_{1}'\tau+\tau^{*}a_{2}'\rho)a_{2}$. Similarly, $\chi_{21} =(\rho^{*}a_{2}'\tau +\tau^{*}a_{1}'\rho)a_{1}$ and $\chi_{22} =(\rho^{*}a_{2}'\rho+\tau^{*}a_{1}'\tau)a_{2}$. Actually, the hole can have multiple scatterings which are the repeats of the basic loops. The total scattering amplitude $r_{j}t^{*}$, representing an Andreev reflected hole in channel $j$ scattered back into channel 0, should satisfy the self-consistent equations
\begin{equation}\label{selfr}
r_{j}=1+\chi_{j1}r_{1}+\chi_{j2}r_{2}.
\end{equation}
The total AR amplitude can be written as $\tilde{a}=(a_{1}r_{1}+a_{2}r_{2})T$, where $T=tt^{*}$ is contributed by the initial electron transmission and the final hole transmission across the Y-junction. With the help of the self-consistent equation (\ref{selfr}), we find
\begin{equation}\label{ra}
\tilde{a}=T[a_{1}+a_{2}-a_{1}a_{2}(a'_{1}+a'_{2})]/(1+\Lambda),
\end{equation}
where $\Lambda=a_{1}a'_{1}a_{2}a'_{2}(1-2T)-(a_{1}a'_{1}+a_{2}a'_{2})|\rho|^{2}-(a_{1}a'_{2}+a'_{1}a_{2})|\tau|^{2}$. In Eq. (\ref{ra}), the condition $\rho\tau^{*}+\rho^{*}\tau=-T$ is utilized, which is obliged by the unitarity of the scattering matrix at the Y-junction. Interestingly, we find that, although $t$, $\tau$, and $\rho$ are all complex numbers, their phases are precisely canceled out and $\tilde{a}$ is determined only by the macroscopic phases of the superconductor.

Experimentally, the AR can be measured by the differential conductance at low temperature. Utilizing the BTK formula \cite{Blonder},
the subgap conductance ($eV<\Delta-|\alpha|$) is twice as much as the AR probability, and is given by
\begin{equation}
\label{RA}
\frac{G}{G_{0}}=\frac{2T^{2}(1+\cos\Phi)(1-\cos\zeta)}{T^{2}\sin^2\zeta+[(1-\cos\zeta)(1-T)+|\tau|^2(\cos\Phi-1)]^2},
\end{equation}
where $G_{0} = e^{2}/h$ is the unit conductance, $\zeta=\phi_{e}+\phi_{h}$, and $\phi_{e,h}=\cos^{-1}[(eV\mp\alpha)/\Delta]$.

Eq. (\ref{RA}) is the major finding of this work. The factor $1+\cos\Phi$ in the numerator of Eq. (\ref{RA}) signifies the interference oscillation of the conductance, from which the Cooper pair momentum in FF superconductors can be directly resolved. The scattering paths for electron and hole are always in pairs (see Fig. \ref{fig2}), therefore all the uncontrollable phases during the scattering at the Y-junction cannot affect the conductance, guaranteed by the particle-hole symmetry. This point marks the major advantage of our Andreev interferometer. When $eV=0$ ($\zeta=\pi$) and $\tau\approx 0$ (no transmission between the two channels), the conductance is reduced to $G/G_{0}=T^{2}(1+\cos\Phi)/(1-T)^{2}$, which is the standard result for an Aharonov-Bohm interferometer, as expected from our basic idea in Fig. 1(b).

\begin{figure}
\centering
\includegraphics[width=0.8\textwidth]{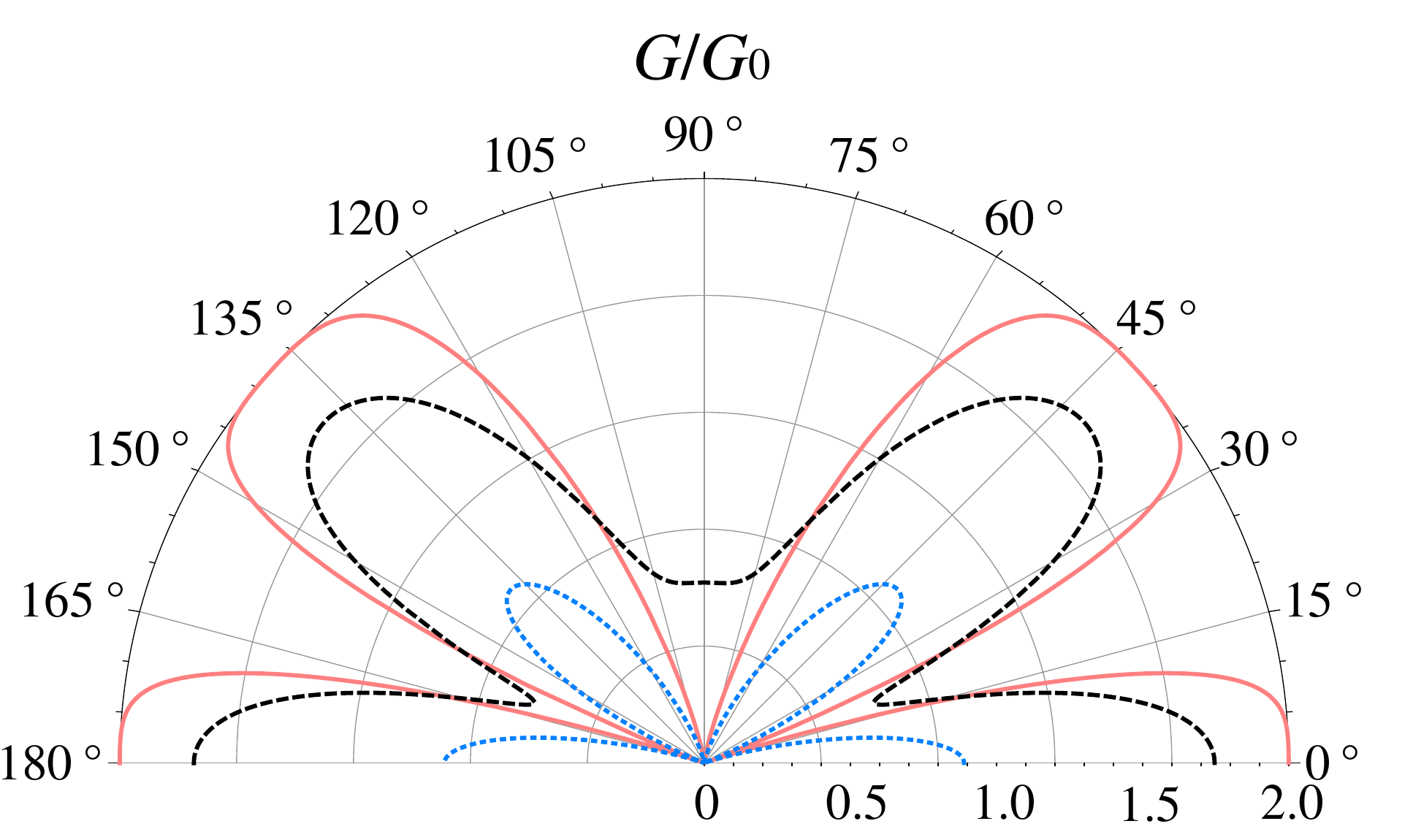}
\caption{(color online) Zero bias conductance as a function of $\theta$. The red solid, blue dotted and black dashed lines are corresponding to the barrier strengths of $Z_{1}=Z_{2}=0$, $Z_{1}=Z_{2}=0.5$ and $Z_{1}=0$, $Z_{2}=0.5$, respectively. See text for values of other parameters}\label{fig3}
\end{figure}

For the general NS interface with a finite barrier, the conductance can be obtained numerically by combining the scattering matrices at the Y-junction and the NS interfaces in a standard way (See appendix 3 for details). In order to present the numerical results, the Cooper pair momentum is expressed in polar coordinates as ${\bm Q} = 2\kappa\xi_{0}^{-1}(\cos\theta,\sin\theta)$, with $\xi_{0}=\hbar v_{F}/\Delta_{0}$ being the superconducting coherence length. The direction angle $\theta$ of the Cooper pair momentum can be precisely determined by carefully tuning the direction of the external Zeeman field \cite{ZZ2}. In the following, we adopt the parameters as $\kappa=0.5, \eta_{1}=\eta_{2}=4, h=3\pi \xi_0$, and $t/\sqrt{2}=\tau=-\rho=0.5$, where $\eta_{j}$ is the length of channel $j$ in unit of $k_{F}^{-1}$, to illustrate our major results.

The zero bias conductance $G/G_{0}$ as a function of $\theta$ is plotted in Fig. \ref{fig3}. We find that the positions for the minimal conductance are irrelevant to the barriers, which always occur at $1+\cos\Phi=0$. For the symmetric NS interfaces ($Z_{1} = Z_{2}$) the minimal conductance reaches zero, while for the unbalanced NS interfaces, it is lifted from zero. Given $-\pi /2<\theta_{1}<\theta_{2}<\pi/2$ being two adjacent minimums, the magnitude of the Cooper pair momentum is given by $2\pi /[h(\sin\theta_{2}-\sin\theta_{1})]$.

\begin{figure}
\centering
\includegraphics[width=0.8\textwidth]{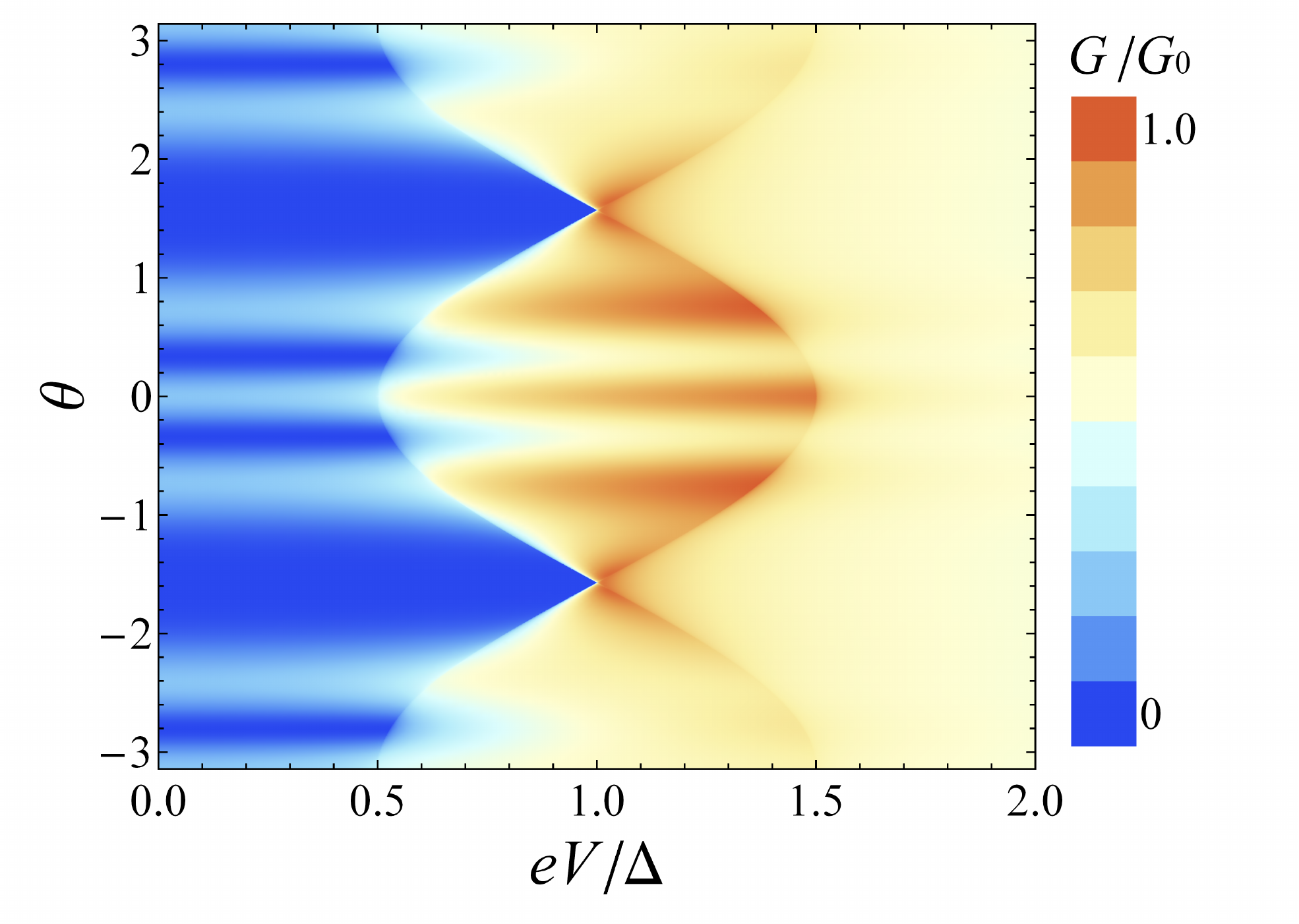}
\caption{(color online) Conductance as a function of $eV$ and $\theta$. $Z_{1}=Z_{2}=1$, and see text for values of other parameters.}
\label{fig4}
\end{figure}

At finite bias, the contour plot of $G/G_{0}$ in $eV$ and $\theta$ plane is shown in Fig. \ref{fig4}. We find that the conductance oscillation with $\theta$ still exists. Moreover, the minimal conductance is insensitive to the bias when $eV < \Delta - |\alpha|$, which greatly relaxes the condition to determine the Cooper pair momentum in the proposed interferometer. In addition, there is a notable boundary in Fig. \ref{fig4} described by $eV/\Delta =1\pm\kappa\cos\theta$ representing the conductance singularities, which is the evidence of the energy spectra split caused by $Q_{x}$.

\section{Discussions}

It is worthwhile to compare the present Andreev interferometer with the conventional ones \cite{Lambert,Samuelsson}. In a conventional interferometer, the conductance oscillation is driven by a supercurrent \cite{Nakano} or an external flux \cite{Samuelsson}, while in our proposal, the conductance oscillates with the direction of the pairing momentum. The momentum direction can be tuned by an in-plane Zeeman field, which does not introduce any orbital effects or fluxes for a two dimensional FF superconductor. In the conventional interferometers, the Zeeman splitting of the conductance is independent of the direction of the magnetic field, while in the present case, the conductance splitting is caused by $Q_{x}$ and strongly depends on the direction angle $\theta$ of the in-plane Zeeman field. As a result, the conductance structure in the present model is specific to the FF states and can be easily distinguished from that in conventional interferometers.

Several important issues related to the experimental implementation of the interferometer should be discussed. The diameter of the nanowires should be smaller than the superconducting coherence length in order to eliminate the superconducting phase fluctuations in the one dimensional channels. For superconductors with a large coherence length ($\xi_{0} \sim 100$ nm in CePt$_3$Si \cite{Frigeri}), the Y-junction may be routinely fabricated using the lithography method. However, for superconductors with a small coherence length ($\xi_{0}\sim 9$ nm \cite{Lortz}), the junction can be realized by carbon nanotubes or other metallic nanowires using the state-of-the-art chemical growing method \cite{YJ1, YJ2,YJ3,YJ4,YJ5,YJ6}. For such nanowires, the transverse modes have an energy spacing about 10 - 100 meV, therefore the electron and hole can transport within a single transverse mode in the energy scale of $\Delta \sim 1$ Kelvin. To observe at least one period of the subgap conductance oscillation, $h \geq \pi\xi_{0}/(2\kappa)$ is required, which can be satisfied in the practical experiments \cite{YJ2,YJ6}. The necessary proximity effects and the associated AR in nanowires have already been demonstrated experimentally by different groups\cite{PX1, PX2, PX3}. To ensure that the phase informations are not lost during the multiple scatterings, the nanowires need a large phase coherence length $L_{\varphi}$. In carbon nanotubes $L_{\varphi} \sim$ 250 nm \cite{Tsukagoshi} and in other semiconductor or metallic nanowires with high mobilities, $L_{\varphi} \sim 1 - 10$ $\mu$m, which is sufficient for our Andreev interferometer.

We need to point out that the proposed interferometer can resolve the FF superconductors with even small pairing momentum, which may not be distinguished by the conventional measurements on their anomalous properties. This interferometer opens the exciting perspectives to detect the Cooper pair momentum in non-centrosymmetric superconductors, which not only can facilitate the unambiguous identification of finite momentum pairing in inhomogeneous superconductors, but also greatly deepen our understanding of fermionic pairing in modern physics.

\section*{Acknowledgements}
We thank Q. H. Cui, S. Raymond, A. Akbari and J. Kaczmarczyk for valuable communications. W. C, R. S, and D. X are supported by 973 Program (Grants No. 2011CB922100, No. 2011CBA00205 and No. 2013CB921804), by NSFC (Grants No. 11074111, No. 11174125, and No. 11023002), by PAPD of Jiangsu Higher Education Institutions, by NCET, and by the Fundamental Research Funds for the Central Universities. M. G is supported by Hong Kong RGC/GRF Projects (No. 401011 and No. 2130352) and the Chinese University of Hong Kong (CUHK) Focused Investments Scheme.

\renewcommand{\theequation}{A.\arabic{equation}}
\setcounter{equation}{0}

\section*{Appendix 1: Microscopic model for proximity effect}\label{a}

We employ a tunneling model to calculate the effective pair potential in the nanowire induced by the two dimensional FF superconductor deposited above. The nanowire is set along $x$-direction, with $y$ being its lateral location. The Hamiltonian of the whole system is composed of three terms, $H=H_{S}+H_{N}+H_{T}$, where $H_{S}$ and $H_{N}$ describe the Hamiltonian in the FF superconductor and that in the nanowire, respectively, and $H_{T}$ describes the tunneling between them. We have
\begin{eqnarray}\label{SMH}
H_{S}=\sum_{\bm{k},\sigma} \varepsilon(\bm{k})c^\dag_{\bm{k} ,\sigma}c_{\bm{k},\sigma}+(\Delta c^\dag_{\bm{k}+\bm{Q}/2,\uparrow}c^\dag_{-\bm{k}+\bm{Q}/2,\downarrow}+{H.c.}), \nonumber \\
H_{N}=\sum_{k_{x},\sigma}\xi(k_{x})b^\dag_{k_{x},\sigma}b_{k_{x},\sigma},\nonumber \\
H_{T}=\Gamma\sum_{\bm{k},\sigma}e^{-ik_{y}y}c^\dag_{k_{x},\sigma}(k_{y})b_{k_{x},\sigma}+{H.c.},
\end{eqnarray}
where $c_{\bm{k},\sigma}$ and $b_{k_{x},\sigma}$ are electron operators for the FF superconductor and the nanowire, respectively, the single-particle energy $\varepsilon(\bm{k})$ and $\xi(k_{x})$ are both measured from the Fermi level, $\Delta$ is the pair potential in the superconductor and $\Gamma$ is the tunneling amplitude between the superconductor and the nanowire. The energy dependence of $\Gamma$ is neglected in the energy scale of $\Delta$, which is much smaller than the Fermi energy. A clean contact between the superconductor and the nanowire (without any scattering caused by impurities) is assumed, so that the momentum component along the nanowire ($k_{x}$) is conserved during the tunneling processes. The pairing in the superconductor occurs with a finite Cooper pair momentum $\bm{Q}$.

The self-energy $\Sigma_N(\omega)$ for the nanowire is given by
\begin{equation}\label{se}
\Sigma_{N}(\omega)=\hat{T}^{\dag}G_{S}(\omega)\hat{T},
\end{equation}
where $G_{S}(\omega)$ is the Green function in the FF superconductor and $\hat{T}$ represents the tunneling terms from the nanowire to the superconductor in $H_{T}$. Given that the Cooper pair momentum is much smaller than the Fermi wave vector of the superconductor $k_{F}^{S}$, the Green function for the FF superconductor can be expressed in the Nambu representation $(c_{\bm{k}+\bm{Q}/2,\uparrow},c^\dag_{-\bm{k}+\bm{Q}/2,\downarrow})$ as
\begin{equation}\label{gf}
G_{S}(\bm{k};\omega)=\frac{\omega-\frac{\hbar^2\bm{k}\cdot \bm{Q}}{2m}+\varepsilon(\bm{k})\tau_z+\Delta\tau_x}{(\omega-\frac{\hbar^2\bm{k}\cdot\bm{Q}}{2m})^2-\varepsilon^2(\bm{k})-\Delta^2},
\end{equation}
where $\tau_{x,y,z}$ are the Pauli matrices in the Nambu space. Inserting Eq. (\ref{gf}) into Eq. (\ref{se}), one obtains the self energy of the nanowire under the basis $(b_{k_{x}+Q_{x}/2,\uparrow},b^\dag_{-k_{x}+Q_{x}/2,\downarrow})$ as
\begin{equation}
\label{se2}
\Sigma_{N}(k_x;\omega)=\\
\Gamma^{2}\sum_{k_y}
\left(
  \begin{array}{cc}
    G_{S}^{11}(\bm{k};\omega) & -e^{iQ_{y}y}G_{S}^{12}(\bm{k};\omega) \\
-e^{-iQ_{y}y}G_{S}^{21}(\bm{k};\omega) & G_{S}^{22}(\bm{k};\omega) \\
  \end{array}
\right)
\end{equation}
Two important conclusions can be drawn from Eq. (\ref{se2}): (a) The proximity effect leads to the finite momentum pairing in the nanowire with $Q_{x}$ equal to that in the FF superconductor; (b) The information of $Q_{y}$ is registered by the phase $e^{\pm iQ_{y}y}$, which defines the macroscopic phases of two channels of the proposed Andreev interferometer.

The electron density in the nanowire is much lower than that in the superconductor so that the $k_{x}$-dependence in Eq. (\ref{se2}) can be dropped. The summation in Eq. (\ref{se2}) can be transformed into a integral over energy $\varepsilon$. The density of states $N(\varepsilon)=[\partial\varepsilon/\partial k_y]^{-1}$ is approximated by its value at the Fermi energy $N(0)$, and one obtains,
\begin{equation}
\Sigma_{N}(\omega)=\sum_{\lambda=\pm}\chi_{\lambda}(\omega)[-\omega_{\lambda}+\Delta(\cos\beta\tau_{x}-\sin\beta\tau_{y})],
\end{equation}
where $\chi_{\pm}(\omega)=\pi N(0)\Gamma^{2}(\Delta^{2}-\omega^{2}_{\pm})^{-1/2}$, $\beta=Q_{y}y$, and $\omega_{\pm}=\omega\pm\hbar^{2}k_{F}^{S}Q_{y}/(2m)$. The Green function of the nanowire is given by,
\begin{equation}
\label{gn}
G_{N}(\omega)=\frac{\mathcal{Z}}{\omega-\mathcal{H}^{eff}_{N}},
\end{equation}
where $\mathcal{H}^{eff}_{N} = \mathcal{Z}[\mathcal{H}_{N}+(\chi_{-}-\chi_{+})\hbar^{2}k_{F}^{S}Q_{y}/(2m)] + (1-\mathcal{Z})\Delta(\cos\beta\tau_{x}-\sin\beta\tau_{y})$, $\mathcal{Z}(\omega)=[1+\chi_{+}(\omega)+\chi_{-}(\omega)]^{-1}$, and $\mathcal{H}_{N}$ is the Hamiltonian $H_{N}$ in Eq. (\ref{SMH}) written in the Nambu space.

Since the Green function is given under the basis of $(b_{k_{x}+Q_{x}/2,\uparrow},b^\dag_{-k_{x}+Q_{x}/2,\downarrow})$, after a Fourier transformation, one finds that the effective pair potential in the nanowire is $\Delta(y)=(1-\mathcal{Z})\Delta e^{iQ_{y}y}e^{iQ_{x}x}$, with the phase modulation being the same as that in the bulk FF superconductor and the strength renormalized by the factor $(1-\mathcal{Z})$. Under the strong coupling limit $\mathcal{Z}\ll 1$, we have $\Delta(y)=\Delta e^{iQ_{y}y}e^{iQ_{x}x}$. The normal part of the Hamiltonian of the nanowire is also modified by the factor $\mathcal{Z}$, indicating a renormalization of the Fermi velocity. Although such a renormalization may result in a velocity mismatch at the NS interface, it can be compensated by utilizing a gate voltage or simulated by a barrier potential. Thus, the effective Hamiltonian in the spatial space for the nanowire can be written as
\begin{equation}\label{SHeff}
H_{eff}=\int dx\sum_{\sigma=\uparrow,\downarrow} \left[\psi^\dag_{\sigma}h_{j}\psi_{\sigma}+(\Delta_{j}\psi^\dag_{\uparrow}\psi^\dag_{\downarrow}
+{H.c.})\right]
\end{equation}
with $h_{j}$ and $\Delta_{j}$ defined in the main text. From Eq. (\ref{SHeff}), the BdG equation in the main text is restored.

\section*{Appendix 2: Blonder-Tinkham-Klapwijk theory}\label{b}

After the transformation $u_{j}=\tilde{u}_{j}e^{i(k+Q_{x}/2)x}$ and $v_{j}=\tilde{v}_{j}e^{i(k-Q_{x}/2)x}$, the BdG equation is reduced to
\begin{equation}\label{BdG_R}
\left(
  \begin{array}{cc}
    \xi_{k+Q_{x}/2} & \Delta e^{iQ_{y}y_{j}} \\
\Delta e^{-iQ_{y}y_{j}} & -\xi_{k-Q_{x}/2} \\
  \end{array}
\right)\left(
         \begin{array}{c}
           \tilde{u}_{j} \\
           \tilde{v}_{j} \\
         \end{array}
       \right)
=E\left(
         \begin{array}{c}
           \tilde{u}_{j} \\
           \tilde{v}_{j} \\
         \end{array}
       \right),
\end{equation}
where $\xi_{k} =\hbar^{2}k^{2}/(2m) - \mu$. The eigenvalues of the BdG equation are found as
\begin{equation}
E =\frac{\xi_{k+Q_{x}/2}-\xi_{k-Q_{x}/2}}{2}\pm\sqrt{\Delta^{2}+\left(\frac{\xi_{k+Q_{x}/2}+\xi_{k-Q_{x}/2}}{2}\right)^2}.
\end{equation}
Under the general condition $Q_{x}\ll k_{F}$, the dispersion Eq. (2) in the main text for the particles near the Fermi level ($|k-k_{F}|\ll k_{F}$) is obtained.

For an electron incident from the normal region with an energy $E$, the wave functions take the form
\begin{eqnarray}\label{state}
\Psi_{N}^{j}=
\left(
  \begin{array}{c}
    1 \\
    a_j \\
  \end{array}
\right)
e^{ik_{F}x}+b_{j}\left(
  \begin{array}{c}
    1 \\
    0 \\
  \end{array}
\right)e^{-ik_{F}x}, \nonumber \\
\Psi_{S}^{j}=
c_{j}
\left(
  \begin{array}{c}
    e^{i\varphi_{j}}u_{0}^{+} \\
    v_{0}^{+} \\
  \end{array}
\right)e^{ik_{F}x}+d_{j}
\left(
  \begin{array}{c}
    e^{i\varphi_{j}}v_{0}^{-} \\
    u_{0}^{-} \\
  \end{array}
\right)e^{-ik_{F}x},
\end{eqnarray}
where $u_{0}^{\pm}=\sqrt{[1+\sqrt{(E\mp\alpha)^{2}-\Delta^{2}}/(E\mp\alpha)]/2}$, $v_{0}^{\pm}=\sqrt{1-(u_{0}^{\pm})^{2}}$, phase $\varphi_{j}=Q_{y}y_{j}$, and all the wave vectors are approximated by $k_{F}$. All the scattering states in Eq. (\ref{state}) are illustrated in Fig. 1(c) in the main text. The amplitudes $a_{j}, b_{j}, c_{j}$ and $d_{j}$ denote the amplitudes of the AR, the normal reflection, and the electron-like and hole-like quasiparticle transmissions, respectively.

Utilizing the boundary conditions $\Psi_{N}^{j}=\Psi_{S}^{j}$ and $\partial_{x}(\Psi_{S}^{j}-\Psi_{N}^{j})=(2mU_{j}/\hbar^{2})\Psi_{N}^{j}$ at $x=L$, the AR and the normal reflection amplitudes are obtained as
\begin{equation}
a_{j}=u_{0}^{-}v_{0}^{+}e^{-i\varphi_j}/\gamma_{j}, \\
b_{j}=-e^{2i\eta_{j}}(u_{0}^{+}u_{0}^{-}-v_{0}^{+}v_{0}^{-})Z_{j}(i+Z_{j})/\gamma_{j},
\end{equation}
where $\gamma_{j}=u_{0}^{+}u_{0}^{-}(1+Z_{j}^{2})-v_{0}^{+}v_{0}^{-}Z_{j}^{2}$, $\eta_{j}$ is the length of channel $j$ in unit of $k_{F}^{-1}$, and $Z_{j}=mU_{j}/(\hbar^{2}k_{F})$ is the dimensionless barrier strength. Similarly, the scattering amplitudes for a incident hole are obtained as
\begin{equation}
a_{j}'=u_{0}^{+}v_{0}^{-}e^{i\varphi_{j}}/\gamma_{j}, \\
b_{j}'=-e^{-2i\eta_{j}}(u_{0}^{+}u_{0}^{-}-v_{0}^{+}v_{0}^{-})Z_{j}(-i+Z_{j})/\gamma_{j}.
\end{equation}

\section*{Appendix 3: Scattering matrix approach}\label{c}

It is convenient to write all the scattering coefficients at the NS interfaces into a reflection matrix as $R_{A}=\left(
                                                                                                              \begin{array}{cc}
                                                                                                                R_{1}&0 \\
                                                                                                                0&R_{2} \\
                                                                                                              \end{array}
                                                                                                            \right)
$, with $R_{j}=\left(
                 \begin{array}{cc}
                   b_{j}&a_{j}' \\
                   a_{j}&b_{j}' \\
                 \end{array}
               \right)
$. The scattering matrix at a symmetric Y-junction takes the form
\begin{equation}S=
\left(
  \begin{array}{cc}
    R_{0}&T_{0}' \\
T_{0}&R_{0}' \\
  \end{array}
\right),
\end{equation}
where the submatrices $T'_{0}=T_{0}^\mathrm{T}=\left(
                                                 \begin{array}{cccc}
                                                   t&0&t&0 \\
                                                   0&t^{*}&0&t^{*} \\
                                                 \end{array}
                                               \right)
$ describe the quasiparticle transmission between chanel 0 and channel $j$, $R_{0}=\sqrt{1-2T}\left(
                                                                                                      \begin{array}{cc}
                                                                                                        1&0 \\
                                                                                                        0&1 \\
                                                                                                      \end{array}
                                                                                                    \right)
$ represents the reflection within channel 0, and $R_{0}'=\left(
                                                            \begin{array}{cccc}
                                                              \rho&0&\tau&0 \\
                                                              0&\rho^{*}&0&\tau^{*} \\
                                                              \tau&0&\rho&0 \\
                                                              0&\tau^{*}&0&\rho^{*} \\
                                                            \end{array}
                                                          \right)
$ represents the reflection within channel $j$ and the transmission between channel 1 and channel 2. The scattering matrix $S$ is written in the particle-hole space, however, all the scattering amplitudes representing the electron-hole conversion are zero at the Y-junction. The scattering amplitudes for electrons are given as $\rho$ for the reflection within channel $j$, $\tau$ for the transmission between channel 1 and channel 2, and $t$ for the transmission between channel 0 and channel $j$, respectively. The scattering amplitudes for holes are just the complex conjugate of those for electrons, due to the particle-hole symmetry of the BdG equation. Due to the time reversal symmetry of the Y-junction, the scattering matrix $S$ is symmetric.

Then, the scattering matrix for the whole Andreev interferometer can be obtained by combining $S$ and $R_{A}$. The unitarity of $S$ gives the restrictions $|\rho-\tau|=1$, $|\rho+\tau|=\sqrt{1-2T}$, and $\rho\tau^{*}+\rho^{*}\tau=-T$, which are useful to simplify the matrix combination. The composite scattering matrix, describing an electron or a hole in channel 0 is reflected back as an electron or a hole, is given by
\begin{equation}
M=R_{0}+T_{0}'R_{A}(1-R_{0}'R_{A})^{-1}T_{0}.
\label{eq-M}
\end{equation}
Specifically, the total AR amplitude in channel 0 for a incident electron is $\tilde{a}=M_{21}$ and the normal reflection amplitude is $\tilde{b}=M_{11}$. The dimensionless conductance is obtained as
\begin{equation}
\frac{G}{G_{0}}=1+|\tilde{a}|^{2}-|\tilde{b}|^{2}.
\end{equation}

\section*{References} %all authors for number<=10; not the title of the paper

\end{document}